\documentclass[abstract=true, DIV=14, parskip=half]{scrartcl}

\usepackage[noEnd,commentColor=black]{algpseudocodex}
\usepackage{amsmath}
\usepackage{amssymb}
\usepackage{amsthm}
\usepackage[mathscr]{euscript}
\usepackage[shortlabels]{enumitem}
\usepackage{graphicx} 
\usepackage{subcaption}
\usepackage{thmtools}
\usepackage{todonotes}
\usepackage{xcolor}
\usepackage{microtype}

\makeatletter
\def\@fnsymbol#1{\ensuremath{\ifcase#1\or \!\;\or \!\;\or \ddagger\or
   \mathsection\or \mathparagraph\or \|\or **\or \dagger\dagger
   \or \ddagger\ddagger \else\@ctrerr\fi}}
\makeatother

\usepackage[normalem]{ulem}
\usepackage{tcolorbox}

\usepackage{upgreek}
\usepackage[colorlinks]{hyperref}
\usepackage[capitalize]{cleveref}

\crefname{equation}{eq.}{eqs.} 
\crefname{enumi}{}{} 
\crefname{icase}{case}{cases}
\crefname{ipart}{part}{parts}
\crefname{iprop}{property}{properties}
\crefname{iinv}{invariant}{invariants}

\crefformat{section}{\S\,#2#1#3}
\crefformat{subsection}{\S\,#2#1#3}
\crefrangeformat{section}{\S\S\,#3#1#4--#5#2#6}
\crefmultiformat{section}{\S\S\,#2#1#3}{,~#2#1#3}{,~#2#1#3}{,~#2#1#3}

\usepackage{authblk}
\usepackage[normalem]{ulem}

\usepackage{tikz}
\usetikzlibrary{calc}

\newcommand{\cF}{\mathcal{F}}

\newcommand{\scs}{\operatorname{scs}}

\DeclareMathOperator{\OPT}{\mathsf{OPT}}
\DeclareMathOperator{\relx}{\mathsf{relax}}

\renewcommand{\alpha}{\upalpha}

\newcommand{\ceil}[1]{\lceil #1 \rceil}

\newcommand{\connected}[1]{\def\temp{#1}\ifx\temp\empty\sim\else\overset{#1}{\sim}\fi}

\newtheorem{theorem}{Theorem}[section]
\newtheorem{lemma}[theorem]{Lemma}

\theoremstyle{definition}

\author[1]{Jialu Hu}
\author[2]{L\'{a}szl\'{o} Kozma}
\affil[1]{TU Munich, Germany \authorcr \texttt{stella.hu@tum.de}}

\affil[2]{TU Dresden, Germany \authorcr \texttt{laszlo.kozma@tu-dresden.de}}

\title{Non-adaptive Bellman-Ford:\\ Yen's improvement is optimal\thanks{$^{*}$Supported by DFG Grant KO 6140/1-2. Work largely done while the authors were at Freie Universität Berlin.}}

\date{}

\begin{document}

\maketitle

\vspace{-0.1in}
\begin{abstract}
The Bellman-Ford algorithm for single-source shortest paths repeatedly updates tentative distances in an operation called \emph{relaxing an edge}. In several important applications a \emph{non-adaptive} (oblivious) implementation is preferred, which means fixing the entire sequence of relaxations upfront, independently of the edge-weights. The original implementation of the algorithm performs, in a dense graph on $n$ vertices, $(1+o(1))n^3 $ relaxations. An improvement by Yen from 1970 reduces the number of relaxations by a factor of two. We show that no further constant-factor improvements are possible, and every \emph{non-adaptive deterministic} algorithm based on relaxations must perform $(\frac{1}{2} - o(1))n^3$ steps. This improves an earlier lower bound of Eppstein of $(\frac{1}{6} - o(1))n^3$. Given that a \emph{non-adaptive randomized} variant of Bellman-Ford with at most $(\frac{1}{3} + o(1))n^3$ relaxations (with high probability) is known, our result implies a strict separation between deterministic and randomized strategies, answering an open question of Eppstein.  

We also address the complexity of finding \emph{short} relaxation sequences for a given input graph on $n$ vertices, answering a question of Eppstein. We show that the problem is co-NP-hard, and moreover essentially inapproximable: While an $n$-approximation is easily obtained, for every $\varepsilon > 0$, no polynomial-time $n^{1-\varepsilon}$-approximation exists, unless P = NP. We further show that \emph{deciding} whether a given relaxation sequence is valid is co-NP-complete, even when the input is the complete graph.

\end{abstract}

\section{Introduction}
The Bellman-Ford algorithm~\cite{Bellman, Ford, Moore} is a classical method for computing \emph{single-source shortest paths} in edge-weighted, directed graphs. Similarly to Dijkstra's algorithm, the Bellman-Ford algorithm uses \emph{edge relaxations}. Relaxing an edge $(u,v)$ means setting the tentative distance from the source $s$ to $v$ to the tentative distance from $s$ to $u$ plus the weight of the edge $(u,v)$, if this is smaller than the current tentative distance from $s$ to $v$. Initially setting the tentative distance from $s$ to $s$ itself to $0$ and to all other vertices to $+\infty$, the algorithm proceeds via repeated edge-relaxations, updating tentative distances until they all reach their final values, the true distances. Relaxing an edge $(u,v)$ changes the tentative distance from $s$ to $v$ to its final (correct) value, if and only if the tentative distance from $s$ to $u$ is already correct and $(u,v)$ is the last edge of a shortest path from $s$ to $v$. Algorithms of this type are also known as \emph{label-correcting}.

Besides the correct handling of negative edge-weights, an important feature of the Bellman-Ford algorithm is that it can be executed in a \emph{non-adaptive} (also called \emph{oblivious}) way. This means that the sequence of edge-relaxations is fixed in advance, independently of the edge-weights of the input graph and of the outcomes of previous operations. This reduces overhead while maintaining the same asymptotic worst-case running time. A non-adaptive sequence of edge-relaxations also makes the algorithm amenable for implementation in parallel or distributed settings, e.g., in network routing protocols~\cite{rfc1058, routing}, where centralized control may not be available.  

It is easy to see that a sequence of (non-adaptive) edge-relaxations correctly computes the distance from $s$ to $v$ if and only if the edges of some shortest path from $s$ to $v$ are relaxed in the order in which they appear on the path (possibly with other edge-relaxations in between). If for every simple path from $s$ to $v$, the edges of the path are relaxed in the order they appear going from $s$ to $v$, then the correct distance from $s$ to $v$ must have been computed. Conversely, if the edges of some simple $s$-to-$v$ path do not appear as a subsequence of the relaxation sequence, then, for some configuration of the edge-weights (e.g., with weights on the path set to $0$ and all other weights set to $1$) the correct distance will not have been reached. 

The problem of finding the most efficient edge-relaxation sequence thus maps to the combinatorial problem of finding the shortest sequence of edges in which every simple path from $s$ to another vertex appears as a subsequence. The standard non-adaptive version of the Bellman-Ford algorithm relaxes all $m$  edges in a round-robin fashion $n-1$ times, where $n$ is the number of vertices. The resulting sequence of edges clearly contains every simple path, and in a complete directed graph the number of edge-relaxations is $(1+o(1))n^3$, with a proportional overall running time.    

A classical improvement by Yen~\cite{Yen} reduces the number of relaxations by a factor of two, while preserving the non-adaptive nature of the algorithm. Yen's algorithm decomposes the edge set of the input graph into \emph{forward edges} and \emph{backward edges}, both sets forming acyclic subgraphs. It then alternates between relaxing all forward- and all backward edges in topological order. 

The question of whether a better relaxation order exists than the one given by Yen's method has been open since 1970. Recently, Eppstein~\cite{Eppstein} showed that every non-adaptive deterministic algorithm must perform $(\frac{1}{6} - o(1))n^3$ relaxations in dense graphs, leaving open the possibility of a further factor-three improvement.
Bannister and Eppstein~\cite{Bannister} obtained a \emph{randomized} improvement to Yen's algorithm, with a high-probability guarantee on the algorithm's success. In the non-adaptive setting, this improves Yen's bound by a factor of $\frac{2}{3}$.

\paragraph{Our results.} We show (\S\,\ref{sec2}) that every non-adaptive deterministic algorithm based on edge-relaxations must perform $(\frac{1}{2} - o(1))n^3$ relaxation steps on a complete directed graph with $n$ vertices. This shows that Yen's algorithm cannot be improved with the choice of a better relaxation sequence, by any constant factor. Moreover, since faster randomized approaches are known, randomization strictly helps for non-adaptive relaxation sequences. This answers an open question of Eppstein~\cite{Eppstein}.

On the complexity side, two natural questions arise. Given a directed graph $G$ and a source vertex $s$, we may ask for a \emph{shortest} non-adaptive sequence of edge-relaxations that result in the correct distances from $s$, for all possible configurations of edge-weights. In light of the previous discussion, this amounts to finding a shortest sequence of edges that contains as subsequence every simple path from $s$ in $G$. A second question is to decide, whether a \emph{given} sequence $S$ of edges is a valid relaxation sequence, i.e., whether it contains every simple path from $s$ in $G$. 

We note that both problems admit polynomial-time algorithms in the easier case of \emph{BF-orderable} graphs, i.e., graphs in which every edge needs to be relaxed at most once; this results from 1980s work of Mehlhorn and Schmidt~\cite{Mehlhorn_BF}, and Haddad and Sch\"{a}ffer~\cite{haddad1988recognizing}.

In this paper we establish the hardness of both tasks, showing that they are unlikely to be efficiently solvable under standard complexity-assumptions. The task of finding short relaxation sequences is addressed in \S\,\ref{sec4} and \S\,\ref{sec:amp}; our result answers a question explicitly raised by Eppstein~\cite{Eppstein}, showing that the problem is co-NP-hard, as well as hard to approximate with a factor below $n^{1-\varepsilon}$ in a graph with $n$ vertices, for every $\varepsilon > 0$. This bound is essentially tight, since an $n$-approximation can be obtained by an easy modification of the standard Bellman-Ford algorithm. The second task, verification, is shown to be co-NP-complete in \S\,\ref{sec3}. Perhaps surprisingly, this result holds even when the input graph $G$ is the complete graph.

\paragraph{Further related work.}
Due to the fundamental nature of the single-source shortest paths problem in the presence of negative weights, several attempts were made to establish its precise complexity or to improve the long-standing $O(mn)$ runtime of the Bellman-Ford algorithm. We refer to Eppstein~\cite{Eppstein} for a discussion of results in different models and for variants of the problem. Note that complexity bounds are sensitive to the exact modeling assumptions, in particular, they differ between the non-adaptive and adaptive settings. 

When adaptivity is allowed, i.e., the algorithm can decide which edge to relax based on the outcomes of past relaxations (and possibly other calculations involving edge-weights), a recent breakthrough by Fineman~\cite{Fineman} has reduced the complexity to $\tilde{O}(mn^{8/9})$, with further improvements~\cite{huang2025faster} to $\tilde{O}(mn^{4/5})$ and most recently~\cite{HairLi26} to $m^{1 + o(1)}$. These results explain why \emph{lower bounds} for adaptive algorithms have been elusive. Nevertheless, an asymptotic lower bound of $\Omega(n^3)$ has also been shown~\cite{Atalig} for adaptive relaxation-based algorithms, with a certain restricted set of operations; while classical Bellman-Ford-variants fall within these restrictions, the mentioned recent improvements do not. 

Of course, the significant improvements in adaptive bounds imply that Yen's variant of Bellman-Ford is no longer competitive in worst-case asymptotic terms. Nonetheless, its simplicity compared to the more recent algorithms will likely ensure its continued relevance in practice. In our view, {non-adaptivity} is not just a technical restriction, but a natural \emph{model of computation}; we find understanding the optimal algorithmic complexity in this basic model a question of intrinsic interest. 

\section{Deterministic lower bound}\label{sec2}
Let $G$ be the complete directed graph with vertex set $V$ and edge set $E$, with $|V| = n$ and $|E| = n(n-1)$, and let $s \in V$ be the source vertex. Let $w:E \rightarrow \mathbb{R}$ denote the edge-weights. Note that $G$ is simple, containing no loops or multiple edges. 

The Bellman-Ford algorithm and its variants compute the distances from $s$ to every vertex in $V$ by initially setting $d(s,s) = 0$ and $d(s,v) = +\infty$ for all $v \in V \setminus \{s\}$, and repeatedly performing operations $\relx(e)$ for edges $e \in E$. The operation $\relx(e)$ for $e = (u,v)$ sets $d(s,v) \leftarrow \min\{d(s,v), d(s,u)+w(e)\}$. The goal is to execute such steps until $d(s,v)$ is the correct distance from $s$ to $v$ for all $v \in V$. We assume that the graph contains no negative weight cycle (such a cycle can be detected by checking if any relax operation still reduces tentative distances after the algorithm has finished). 

As we consider non-adaptive algorithms, the behaviour of the algorithm is fully specified by the sequence of relax operations $S = (e_1, \dots, e_L)$ it performs, where $e_i \in E$ for all $1 \leq i \leq L$. We refer to $L = |S|$ as the length of the sequence. Our first main result is the following.

\begin{theorem}\label{thm1}
Every valid relaxation sequence for $G$ must have length at least $(\frac{1}{2}-o(1))n^3$.
\end{theorem}

In the remainder of the section we prove the theorem. All paths considered are simple. We say that a sequence $S$ of edges \emph{contains} a path $(v_1, \dots, v_k)$ if $S$ contains $(v_1,v_2), (v_2,v_3), \dots, (v_{k-1},v_k)$ as a subsequence (in the given order, but not necessarily contiguously). A \emph{$k$-path} is a path $(s,v_1,\dots,v_k)$ for some set of $k$ vertices $\{v_1, \dots, v_k\} \subseteq V \setminus \{s\}$. We say that a sequence of edges $S$ is \emph{$k$-valid}, if it contains every possible $k$-path. By our earlier discussion, a sequence $S$ is a valid relaxation sequence for $G$ in a non-adaptive single-source shortest path algorithm, if and only if it is $(n-1)$-valid.  

The task of finding a short $k$-valid sequence appears related to the task of finding a short \emph{permutation supersequence}, a well-studied question in combinatorics. 
In that context, $c_n$ denotes the smallest possible length of a sequence over $\{1,\dots,n\}$ that contains each permutation of $\{1,\dots,n\}$ as a subsequence (not necessarily contiguously).

It is known that $c_n \leq \ceil{n^2-\frac{7}{3}n + \frac{19}{3}}$~\cite{Radomirovic}, see also the survey of Engen and Vatter~\cite{engen2020containing}. Most relevantly for us, a lower bound of the form $c_n \geq (1- o(1))n^2$ was shown in 1976 by Kleitman and Kwiatkowski.

\begin{lemma}[\cite{kleitman}]\label{lem1}
For all $n \geq 1$ and all $\varepsilon > 0$, we have $c_n > n^2 - C_\varepsilon \cdot n^{7/4 + \varepsilon}$, with $C_\varepsilon$ depending only on $\varepsilon$. 
\end{lemma}

We use Lemma~\ref{lem1} to prove Theorem~\ref{thm1}. 

It is well-known that whenever $N$ is even, the edge-set of a complete (undirected) graph on $N$ vertices can be decomposed into $N-1$ disjoint perfect matchings, i.e., sets of $\frac{N}{2}$ non-incident edges. (Such a decomposition is called a \emph{$1$-factorization}, often studied in the context of round-robin tournaments, e.g., see~\cite{mendelsohn1985one}). If $N$ is odd, we can simply ignore one of the vertices and decompose the remainder of the graph into $N-2$ disjoint matchings, each consisting of $\frac{N-1}2$ edges. 

Denote the vertex set of the complete directed graph $G$ as $V = \{s,v_1,\dots,v_{n-1}\}$. 
Consider a complete (undirected) graph on $\{v_1, \dots, v_{n-1}\}$. By the previous paragraph, regardless of the parity of $n$, we can find in this graph $n-3$ disjoint matchings, each of size at least $\lfloor \frac{n-2}{2} \rfloor$. Denote such a collection of matchings as $E_1, \dots, E_{n-3}$ (each $E_i$ is a set of undirected edges).

For each matching $E_i$ we form two sets of \emph{directed} edges in $G$; one where each edge of $E_i$ is oriented from smaller to larger index vertex (call this $F_i$), and one where each edge of $E_i$ is oriented from larger to smaller index vertex (call this ${H_i}$). We thus obtain a collection $\mathcal{C} = \{F_1,H_1, \dots, F_{n-3}, {H_{n-3}}\}$ of pairwise disjoint sets, each consisting of pairwise non-incident directed edges in $G$. 

Let $S$ be a shortest $(n-1)$-valid, or simply valid, sequence of edges for graph $G$. For any $F \in \mathcal{C}$ let $S_F$ denote the maximal subsequence of $S$ consisting of elements in $F$.

We claim that $S_F$ must contain every permutation of $F$ as a subsequence, and thus the lower bound of Lemma~\ref{lem1} applies to $S_F$.

Indeed, suppose some permutation of $F$ is not contained in $S_F$, and denote this permutation (of directed edges of $G$) as $(e_1, \dots, e_{|F|})$, where $e_i$ is the directed edge $(x_i, y_i)$, for $x_i,y_i \in V \setminus \{s\}$. 
In this case, $S$ would not contain the sequence of edges $( (s,x_1), \mathbf{e_1}, (y_1,x_2), \mathbf{e_2}, (y_2,x_3), \dots, \mathbf{e_{|F|}} )$ that forms a simple path from the source to $y_{|F|}$, and thus $S$ would not be valid. 

Since $|F| \ge \lfloor \frac{n-2}{2} \rfloor$ for all $F \in \mathcal{C}$, we have $|S_F| \geq (\frac{1}{4}-o(1))n^2$, by Lemma~\ref{lem1}. As edge sets in $\mathcal{C}$ are pairwise disjoint, the corresponding subsequences of $S$ do not overlap. Thus, we have $$|S| ~\geq~ |\mathcal{C}| \cdot \left(\frac{1}{4}-o(1)\right)n^2 ~\geq~ \left(2n-6\right)\left(\frac{1}{4}-o(1)\right)n^2 ~\geq~ \left(\frac{1}{2}-o(1)\right)n^3. \quad\quad\quad\quad~~~~~~\qed $$

\section{Complexity results}

In this section we show our main complexity-results, for the problem of deciding the validity of a given relaxation sequence (\S\,\ref{sec3}) and for the problem of finding the shortest valid relaxation sequence (\S\,\ref{sec4}). Note that the hardness of either problem does not immediately imply hardness of the other. In \S\,\ref{sec:amp}, building on the construction of \S\,\ref{sec4}, we show an \emph{almost linear} hardness-of-approximation result for the same problem.

\subsection{Deciding validity}\label{sec3}

Let $G$ be a directed graph, with a special vertex $s$. 
Let $S = (e_1, \dots, e_{|S|})$ be a sequence of edges of $G$. The \emph{relaxation validity} question asks whether $S$ contains as subsequence every simple path of $G$ starting in $s$, in other words, whether $S$ is a valid relaxation sequence. 

\begin{theorem}\label{thm2}
    Relaxation validity is co-NP-complete.
\end{theorem}

In the remainder of the subsection we prove Theorem~\ref{thm2}. Membership in co-NP is trivial; if $S$ is not valid, then some simple path $P$ witnesses this. 
Verifying that $S$ does not contain $P$ can be done by greedily matching edges of $P$ in their order from $s$, to the earliest occurrence in $S$, to the right of the entry matched to the predecessor edge. 

To show co-NP-hardness, we reduce from the \emph{all-permutations supersequence} (APS) problem, shown to be co-NP-hard by Uzna\'{n}ski~\cite{Uznanski}.
APS asks whether a given sequence $T = (t_1, \dots, t_{|T|})$ contains every permutation of $\{1,\dots,n\}$ as a subsequence. (Alternatively, one may ask whether there is a permutation \emph{not} contained in $T$, which is consequently NP-hard).

The reduction from APS to relaxation validity is straightforward. Consider a complete directed graph $G$ with a set of $2n+1$ vertices denoted $V = \{s,x_1,y_1, \dots, x_n,y_n\}$. For $i \in \{1,\dots,n\}$, let $e_i$ denote the edge $(x_i, y_i)$. We refer to these edges as \emph{special edges}.  

Let $Q$ be a valid relaxation sequence of $G$, i.e., a $2n$-valid sequence of edges. Let $R$ be the sequence of edges obtained from $Q$ by deleting every occurrence of $e_i$ for all $i$.

We construct a relaxation sequence $S$ of $G$ as follows. For $i = 1,\dots,|T|$, form $S_i$ by appending to the end of sequence $R$ the edge $e_j$, for $j=t_i$. Then, concatenate $S_1, \dots, S_{|T|}$, and finally $R$, to form the sequence $S$. From the preceding discussion, we can enforce $|Q|,|R| \in O(n^3)$ and the construction of $S$ takes time polynomial in $|T|$. Note that we do not require $Q$ to be optimal, a simple round-robin sequence of length $O(n^3)$ is sufficient.

We claim that $T$ contains every permutation of $\{1, \dots, n\}$ if and only if $S$ contains every simple path starting from $s$ in $G$, from which Theorem~\ref{thm2} follows.

First, suppose $T$ does not contain some permutation $\pi_1, \dots, \pi_n$. Then, $S$ does not contain the sequence of edges $(e_{\pi_1}, \dots, e_{\pi_n})$ as these edges only appear in $S$ in correspondence to entries of $T$. Consequently, $S$ cannot correctly relax the path $(s,x_{\pi_1}, y_{\pi_1}, \dots, x_{\pi_n},y_{\pi_n})$.

Suppose now that $T$ contains every permutation of $\{1, \dots, n\}$. Then, every simple $k$-path $P$ is correctly relaxed by $S$. 
Indeed, since $T$ contains every permutation, $S$ contains the (possibly incomplete) permutation of edges $e_{\pi_1}, e_{\pi_2}, \dots$ in $P$, in their order of appearance in $P$. 

The portion of $P$ before the first special edge is a simple path from $s$, free of special edges, hence contained in the first copy of $R$. Any other portion of $P$ between special edges or after the last special edge is a special-edge-free simple path, and thus contained in a respective copy of $R$. 
This concludes the proof.   

\subsection{Shortest relaxation sequence}\label{sec4}

Again, let $G$ be a directed graph, with a special vertex $s$. We study the problem of finding a shortest possible sequence $S$ of edges of $G$ that is a valid relaxation sequence. The decision version of the problem asks, given $k$, whether a valid relaxation sequence of length at most $k$ exists. We refer to this problem as \emph{shortest relaxation sequence} (SRS). In this subsection we prove the hardness of the decision problem, as well as hardness of approximating the solution by a constant factor. The latter can be seen as a warmup towards a stronger result; in \S\,\ref{sec:amp} we amplify the lower bound construction to show hardness of approximation by a factor almost linear in the number of input vertices. 

\begin{theorem}\label{thm3}
    SRS is co-NP-hard. Moreover, for every $\varepsilon > 0$, there is no polynomial-time $(3/2 - \varepsilon)$-approximation algorithm for the shortest valid relaxation sequence, unless P = NP.
\end{theorem}

Note that the co-NP-hardness of the one-sided decision problem implies the NP-hardness of the exact optimization problem (under polynomial-time Turing reductions). Membership in co-NP remains unclear; it would require certifying that \emph{every} valid sequence has length greater than a given bound. The result of \S\,\ref{sec3} also shows that a candidate sequence itself cannot serve as a polynomial-time checkable certificate for the existence of a valid sequence of a certain length, unless P=NP. We note that since candidate sequences have polynomial length (a simple round-robin sequence gives an upper bound of $m(n-1)$), the SRS problem is clearly in $\Sigma_2^P$; its $\Sigma_2^P$-completeness remains an intriguing possibility.

We proceed with the proof of Theorem~\ref{thm3}. Our starting point is the \emph{$2$-disjoint-paths} (2DP) problem. In 2DP, given a directed graph $H$ and four distinct vertices $a,b,c,d$, the task is to decide whether there exist two vertex-disjoint paths in $H$, one going from  $a$ to $b$, and one from $c$ to $d$. A classic result by Fortune, Hopcroft, and Wyllie~\cite{Fortune} shows that 2DP is NP-complete.

\paragraph{Reduction.}
Let ($H,a,b,c,d$) be an instance of 2DP and let $n$ and $m$ denote the number of vertices, resp., edges of $H$. We create an instance $G$ of SRS. Towards this goal, we set $q = (n+1)(m+3)$ and $L=Cq$, for a constant $C$ to be fixed later. 

We build $G$ by adding to $H$: 
\begin{enumerate}[(i)]
\item new vertices $s$, $t$, and edges $(s,a), (s,d)$, and $(t,b)$; recall that $a,b,c,d$ are the special vertices of the 2DP instance;
\item a directed path $A$ of exactly $L$ edges from $b$ to $c$;
\item a directed path $B$ of exactly $L$ edges from $d$ to $t$.
\end{enumerate}

\begin{figure} 
\centering\includegraphics[scale=0.45]{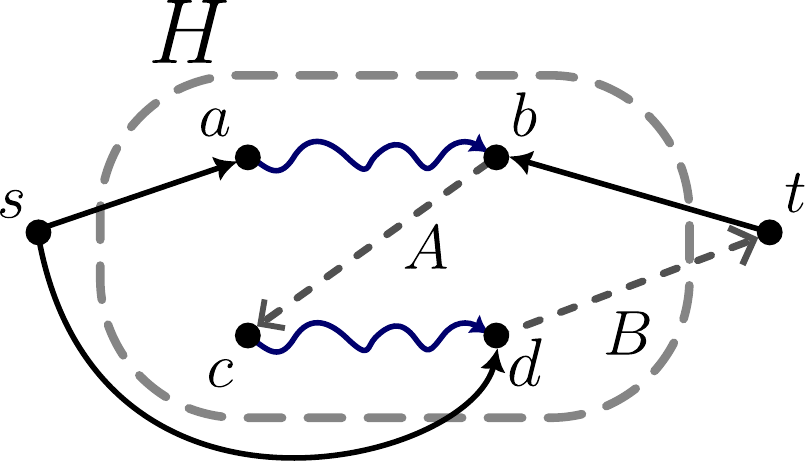}
\captionsetup{width=.92\linewidth}
\caption{Construction of graph $G$ from graph $H$ in the reduction for Theorem~\ref{thm3}. Vertices $s$, $t$, edges $(s,a)$, $(s,d)$, $(t,b)$ and paths $A$ from $b$ to $c$ and $B$ from $d$ to $t$ are new. \label{fig1} }
\end{figure}

All intermediate vertices on the paths $A$ and $B$ are new and distinct. By $A$ and $B$ we also denote the sequences of edges on the respective paths. The construction of $G$ clearly takes polynomial time. See Figure~\ref{fig1} for an illustration. 

We start with a simple characterization of YES-instances for 2DP, which yields a bound on the solution size for SRS.

\begin{lemma}\label{lema}
($H,a,b,c,d)$ is a YES-instance of 2DP if and only if there is a simple path in $G$ starting from $s$ that traverses $A$ and then $B$ (with possibly other edges before, after, or in between).
\end{lemma}

\begin{proof}
For a YES-instance of 2DP we can concatenate (1) the edge $(s,a)$, (2) the path from $a$ to $b$ in $H$, (3) path $A$, (4) the path from $c$ to $d$ in $H$, and (5) path $B$, to obtain the claimed path in $G$. 

Conversely, if a simple path $P$ in $G$ starts from $s$ and traverses $A$ before $B$, then it must start with the edge $(s,a)$. Indeed, starting with $(s,d)$ would visit $d$ before $A$ and traversing $B$ later would necessarily revisit $d$.

Before reaching $A$, the path $P$ cannot visit $t$, since the only incoming edge of $t$ is the final edge of $B$. It follows that the part of $P$ from $a$ to $b$ lies entirely in $H$. Similarly, between leaving $A$ at $c$ and entering $B$ at $d$, path $P$ must stay entirely in $H$. These two portions of $P$ are vertex-disjoint since $P$ itself is simple. It follows that ($H,a,b,c,d$) is a YES-instance. \end{proof}

\begin{lemma}\label{lemb}
If ($H,a,b,c,d$) is a YES-instance, then every valid relaxation sequence of $G$ has length at least $3L = 3Cq$.
\end{lemma}

\begin{proof}
Lemma~\ref{lema} establishes that a valid relaxation sequence for $G$ must contain $AB$ as a subsequence. 
Consider a path $P'$ consisting of concatenating $(s,d)$, $B$, $(t,b)$, $A$. Since $P'$ is a simple path in $G$ starting from $s$, any valid relaxation sequence must also contain $P'$, and consequently $BA$ as a subsequence. Since $A$ and $B$ both consist of $L$ distinct edges and their edge sets are disjoint, any sequence containing both $AB$ and $BA$ as a subsequence must have at least $3L$ elements, since for all $i$ it must contain either $a_i,b_i,a_i$ or $b_i,a_i,b_i$ as a subsequence, where $a_i,b_i$ are the $i$-th elements of $A$, resp., $B$.  
\end{proof}

We now move to the case when ($H,a,b,c,d$) is a NO-instance. Then, no simple path in $G$ starting from $s$ can visit any edge of $A$ and subsequently any edge of $B$. Indeed, if the path enters $A$ or $B$, then it must do so from their beginning. Since the path does not stop inside $A$ (since it later must visit $B$), it must continue until its endpoint. Even if the path stops in $B$ partway, we can extend it to $t$, while maintaining it as a simple path, since none of the remaining vertices of $B$ have been visited earlier. We reach a contradiction, since the path now traverses $A$ and then $B$, which, by Lemma~\ref{lema} means that ($H,a,b,c,d$) is a YES-instance.

\begin{lemma}\label{lemc}
If ($H,a,b,c,d$) is a NO-instance, then every valid relaxation sequence of $G$ has length at most $(2C + 3)q$.
\end{lemma}

\begin{proof}
Let $Q$ be a sequence of the $m+3$ edges other than those in $A$ and $B$, each appearing once, in any order, and let $R$ consist of $Q$ repeated $n+1$ times.

Note that $|R|=q$. Every simple path that is edge-disjoint from $A$ and $B$ is contained in $R$ as a subsequence, since it only contains the original $n$ vertices of $H$ and possibly $s$ and/or $t$ and at most $n+1$ edges. 

Consider the sequence $W = R\,B\,R\,A\,R$. By our earlier observations, and since ($H,a,b,c,d$) is a NO-instance, a simple path starting from $s$ cannot visit any edge of $A$ before any edge of $B$. Moreover, edges of $A$ and edges of $B$ occurring on the path must form contiguous subpaths, and cannot be visited again. The three copies of $R$ cover the subpaths before, between, and after the visits of edges in $B$ and $A$. It follows that $W$ contains every simple path beginning from $s$. Since $|W| = 3q + 2L = (2C+3)q$, the claim follows. \hfill \qedhere
\end{proof}

The reduction is now complete: by Lemmas~\ref{lemb} and \ref{lemc}, if we set $C=4$, then the length of every valid relaxation sequence of $G$ is at least $12q$ in case of a YES-instance of 2DP, and at most $11q$ in case of a NO-instance. An algorithm for SRS can determine whether the second condition holds, and consequently, whether ($H,a,b,c,d$) is a NO-instance. A polynomial-time algorithm of this kind would thus imply P = co-NP. 

To show hardness of approximation, fix $\upalpha < 3/2$ and set $\varepsilon = 3/2 - \upalpha$ (assume, w.l.o.g., $\varepsilon < 1/2$) and $C = \lceil 3/\varepsilon \rceil > 3$. By Lemmas~\ref{lemb} and \ref{lemc}, a YES-instance implies an SRS optimum of at least $3Cq$ and a NO-instance implies an SRS optimum of at most $(2C+3)q$. Since $3Cq/((2C+3)q) > 3/2 - \varepsilon = \upalpha$, an $\upalpha$-approximation returns a value below $3Cq$ on every NO-instance and at least $3Cq$ on every YES-instance, deciding 2DP. Thus, a polynomial-time $\upalpha$-approximation would imply P = NP.

Note that both the number of edges and the number of vertices of $G$ is $\Theta(mn)$, so the hardness results apply even to sparse graphs.

\subsection{Stronger hardness of approximation}\label{sec:amp}

In this section we amplify the reduction of \S\,\ref{sec4} to show our main complexity-result.

\begin{theorem}\label{thm4}
    For every $\varepsilon > 0$, no polynomial-time algorithm exists that approximates the shortest valid relaxation sequence on input graphs on $n$ vertices with a factor $n^{1 - \varepsilon}$, unless P = NP. 
\end{theorem}

\begin{figure} 
\centering\includegraphics[scale=0.55]{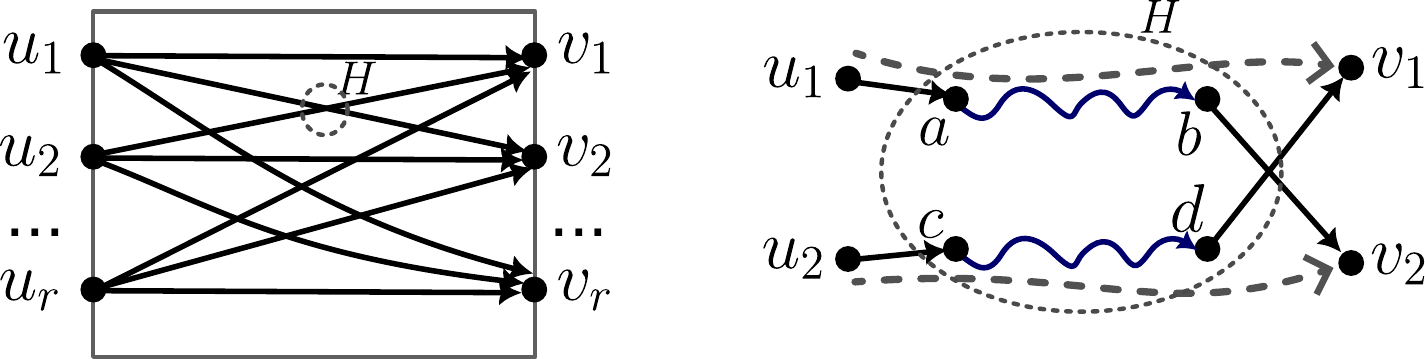}
\captionsetup{width=.92\linewidth}
\caption{(left) Routing gadget, showing the replacement of a crossing between curves $u_1 \rightarrow v_2$ and $u_2 \rightarrow v_1$ by a copy of $H$. (All crossings are similarly replaced); (right) Zooming into the crossing gadget $H$. If $H$ is a YES-instance, then routing along the original curves is possible. If $H$ is a NO-instance, then two paths entering the crossing can only be routed via $a \rightarrow d$ and $c \rightarrow b$ (if at all), resulting in an uncrossing of the original curves, indicated by the dashed arrows. \label{fig2} }
\end{figure}

We use a similar reduction from 2DP to SRS as in \S\,\ref{sec4}, but we recursively compose the graph gadget used earlier. Embedding two recursive subgraph-gadgets instead of the paths $A$ and $B$ would lead to a lower bound factor of $n^{\log_2{(3/2)}} \approx n^{0.585}$. To obtain the stated stronger result, we need to combine a higher number of recursive gadgets, with a \emph{routing network} that ensures that they can be traversed by a simple path in every possible order. 

\paragraph{Routing network.} Let ($H,a,b,c,d$) be an instance of $2DP$, and let $n$ and $m$ denote the number of vertices, resp., edges of $H$. 
We first describe the routing gadget that has $2r$ special vertices called \emph{ports}: $r$ such vertices $u_1, \dots, u_r$ are \emph{entry ports} and $r$ such vertices $v_1, \dots, v_r$ are \emph{exit ports}, for some constant $r \geq 3$ fixed later. 

We construct the gadget such as to ensure that the following holds:

(1) if ($H,a,b,c,d$) is a YES-instance, then every matching between entry and exit ports can be routed through pairwise vertex-disjoint directed paths;

(2) if ($H,a,b,c,d$) is a NO-instance, then disjoint routes preserve port order in the sense that disjoint paths $u_i \rightarrow v_j$ and $u_{i'} \rightarrow v_{j'}$ imply $i<i' \iff j<j'$.

We achieve this by constructing the graph geometrically: we place entry-, resp., exit ports on the vertical sides of an axis-parallel rectangle, ordered by their indices. We then draw all $r^2$ directed wires ($u_i$ to $v_j$) from left to right as Jordan curves, where curves $u_i \rightarrow v_j$ and $u_{i'} \rightarrow v_{j'}$ cross inside the rectangle exactly if $(i,i')$ and $(j,j')$ are ordered differently; multiple crossings and tangencies are not allowed. (There are $\Theta(r^4)$ crossings). Finally, we replace each crossing by an individual copy of the graph $H$, attaching the two incoming wire ends at $a$ and $c$ and their respective continuations at $b$ and $d$, so that the original crossing connections are $a$ to $b$ and $c$ to $d$. All connections between crossing gadgets run left to right. See Figure~\ref{fig2} for illustration.

We can easily verify that the conditions stated above hold. In the YES case, select the wires belonging to the desired matching. At each crossing used by two selected wires, route them using the disjoint $a \rightarrow b$ and $c \rightarrow d$ paths. This realizes the matching disjointly.
For the NO case, consider any collection of vertex-disjoint network routes. A route cannot revisit a crossing gadget, since connections between gadgets run left to right. If two routes use one gadget, the crossing pairing $a \rightarrow b$ and $c \rightarrow d$ is impossible. Their pairing must therefore be the non-crossing one
with $a \rightarrow d$ and $c\rightarrow b$.

If only one route uses a gadget, its connection can also be drawn without intersecting another route there. Consequently, all selected routes can be represented by pairwise disjoint curves across the rectangle, and their terminal order must be preserved, proving (2).

We next describe the recursive construction of a sequence of graphs $G_k$, each of which has two terminal vertices (referred to as \emph{entry} and \emph{exit}). 
As before, we choose a certain path length $L$, which will be determined later. As the base case, $G_0$ is a directed path of $L$ edges from the entry vertex to the exit vertex. 

\begin{figure} 
\centering\includegraphics[scale=0.45]{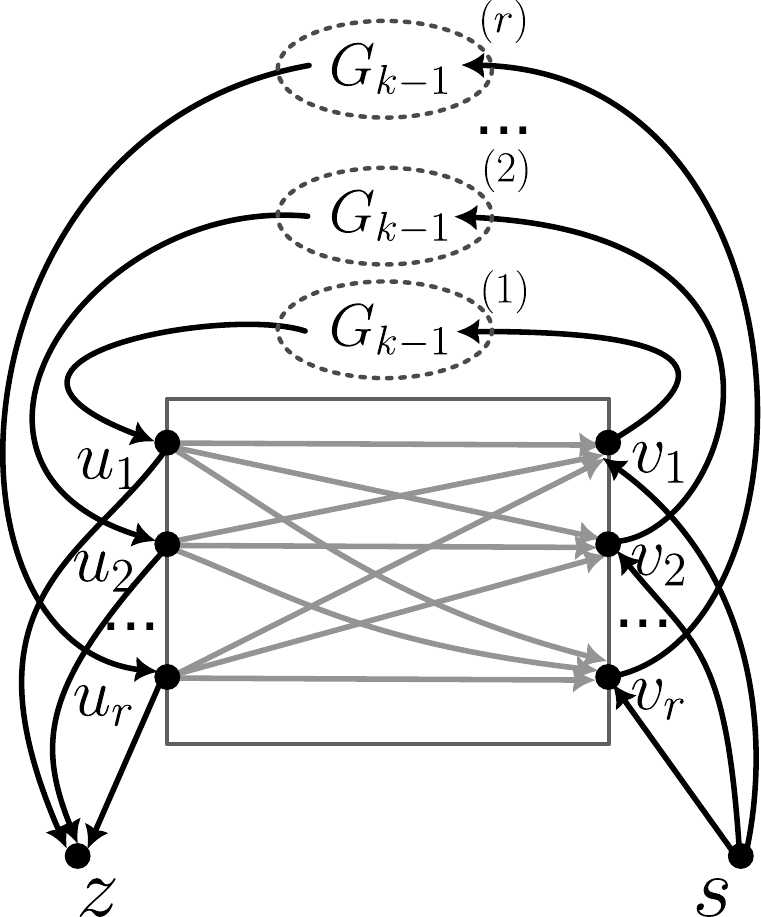}
\captionsetup{width=.92\linewidth}
\caption{Construction of graph $G_k$ in the reduction for Theorem~\ref{thm4}. Routing network gadget shown as rectangle, with exit/entry ports identified with entry, resp., exit vertices of the corresponding copy of $G_{k-1}$; copies of $G_{k-1}$ indexed $1$ to $r$, bottom to top.   \label{fig3} }
\end{figure}

For $G_k$, where $k>0$, take a new copy of the routing gadget and $r$ new copies of $G_{k-1}$. For the $i$-th copy of $G_{k-1}$, identify its entry with $v_i$ and its exit with $u_i$, the exit, resp., entry ports of the routing gadget. Add entry vertex $s$ and exit vertex $z$, with edges $(s,v_i)$, $(u_i,z)$, for all $1\le i\le r$.
See Figure~\ref{fig3} for an illustration. 

Observe that traversing the $i$-th copy of $G_{k-1}$ takes a path from $v_i$ to $u_i$, after which the network can lead to another copy of $G_{k-1}$. 
The interiors of the $G_{k-1}$ instances are otherwise disjoint from each other and from the surrounding graph. The entry vertex of $G_{k-1}$ has no incoming edge from its interior and its exit vertex has no outgoing edge to its interior.\footnote{The construction of $G_k$ is similar in spirit to the construction of $G$ in \S\,\ref{sec4}, with the copies of $G_{k-1}$ playing the role of the paths $A$ and $B$ in the earlier reduction.}

We next bound the length of the shortest valid relaxation sequence in $G_k$ from source $s$, denoted $\OPT_k$.

\paragraph{Case: ($H,a,b,c,d$) is a NO-instance.}

In a NO-instance, every simple path starting from $s$ visits the copies of $G_{k-1}$ in monotone (increasing or decreasing) order of their indices. Suppose the successively visited copies have indices
$i_1,i_2,\ldots,i_t$. These indices are distinct, and between consecutive visits, the path uses the network routes $u_{i_j} \rightarrow v_{i_{j+1}}$. These routes are pairwise vertex-disjoint because they are disjoint portions of one simple path.
Applying (2) to two consecutive routes, we obtain $
i_j<i_{j+1} \quad\Longleftrightarrow\quad i_{j+1}<i_{j+2}$.
Hence the index sequence is entirely increasing or entirely decreasing. Note that the last visit may be incomplete, which does not affect the argument. 

Let $\nu, \mu$ be the numbers of vertices and edges in the graph before inserting the copies of $G_{k-1}$, including $s$, $z$, and their incident edges. These quantities depend on $H$ and $r$, but not on the recursion depth $k$. Call these $\mu$ edges \emph{ordinary}.

Let $Q$ be the sequence of all ordinary edges (in any order), and let $R=Q^{\nu-1}$, and $q=|R|=(\nu-1)\mu$. It is easy to see that every portion of a simple path consisting only of ordinary edges is contained in $R$ as a subsequence. 

Inductively assume that $W_i$ is a valid relaxation sequence for the $i$-th copy of $G_{k-1}$, and consider the sequence consisting of $$W_1,W_2,\ldots,W_r,W_{r-1},\ldots,W_1,$$
with $R$ inserted before and after the sequence, as well as between every two $W$-blocks. This sequence contains as a subsequence both increasing and decreasing visit-orders of the $G_{k-1}$-copies, including paths ending inside a copy or in the network, it is thus a valid relaxation sequence for $G_k$. Let $L_k$ denote its length. 

Writing $A=2r-1$, the resulting sequence lengths satisfy 
$L_0=L$, and $$L_k=A L_{k-1}+2rq.$$

Let us choose $L=2rq$. Then the shortest valid relaxation sequence of $G_k$ has length
\begin{equation}
\OPT_k \le L_k =L\left(A^k+\frac{A^k-1}{A-1}\right) <2A^kL.\label{eqn1}
\end{equation}

Before treating the YES-instances, let us state a useful lemma. 
Let $\scs(\cF)$ denote the length of the shortest common supersequence of a family of sequences $\cF$.

\begin{lemma}\label{lem:help}
Suppose $\mathcal A_1,\ldots,\mathcal A_r$ are families of sequences over pairwise disjoint alphabets, each with shortest common supersequence length at least $p$. Define
$$
\operatorname{Sym}(\mathcal A_1,\ldots,\mathcal A_r)
=
\bigcup_{\pi\in S_r}
\mathcal A_{\pi(1)}\cdots\mathcal A_{\pi(r)}.
$$
Then
$$\scs\!\left(\operatorname{Sym}(\mathcal A_1,\ldots,\mathcal A_r)\right)
\ge \binom{r+1}{2}p.$$

\end{lemma}
\begin{proof}
Consider a sequence $W$ that is a supersequence of every element of this family. Let $t_i$ be the earliest position at which the prefix of $W$ contains all of $\mathcal A_i$. Choose $i$ maximizing $t_i$. The prefix up to $t_i$ contains all $r$ families and therefore has length at least $rp$, because their alphabets are disjoint.
Consider the sequence $w\in\mathcal A_i$ whose earliest completion is $t_i$. Since $W$ must contain $w$ followed by every permutation and every choice of words from the remaining families, the suffix strictly after $t_i$ 
is a common supersequence of $\operatorname{Sym}((A_\ell)_{\ell \neq i})$. 
Induction on $r$ (with the case $r=1$ trivial) gives
$$
\quad\quad\quad\quad\quad\quad\quad\quad\quad\quad\quad\quad  |W|\ge rp+\binom r2p=\binom{r+1}{2}p. \quad\quad\quad\quad\quad\quad\quad\quad\qedhere
$$
\end{proof}

\paragraph{Case: ($H,a,b,c,d$) is a YES-instance.}

In a YES-instance, every permutation of the $G_{k-1}$ instances is possible. For a permutation $i_1,\ldots,i_r$, route
$u_{i_1}\rightarrow v_{i_2}, u_{i_2}\rightarrow v_{i_3}, \dots, u_{i_{r-1}}\rightarrow v_{i_r}$ through the network. These connections form a matching between distinct entry and exit ports.
Combining them with any independently chosen entry-to-exit path in each copy of $G_{k-1}$ yields a simple path
$$
s\to v_{i_1}
\;\xrightarrow{\text{copy }i_1}\;
u_{i_1}\xrightarrow{} v_{i_2}
\;\xrightarrow{\text{copy }i_2}\;\cdots
\xrightarrow{\text{copy }i_r}u_{i_r}\to z.
$$

Restrict the path to the bottom-level path edges (i.e., those appearing in copies of $G_0$), deleting all other edges. We define $\mathcal F_0$ to be the singleton family containing the sequence of edges in the bottom-level path, and let
\[
\mathcal F_{k-1}^{(i)}
=\{\phi_i(w):w\in\mathcal F_{k-1}\},
\]
where \(\phi_i\) relabels each edge by its corresponding edge in the $i$-th copy of $G_{k-1}$ in the graph. Set
$$\mathcal F_k
=
\operatorname{Sym}\bigl(
\mathcal F_{k-1}^{(1)},\ldots,\mathcal F_{k-1}^{(r)}
\bigr).$$

Every sequence in \(\mathcal F_k\) is the projection of a simple entry-to-exit path in $G_k$.
The alphabets used for the $r$ copies of $G_{k-1}$ are disjoint, so Lemma~\ref{lem:help}, with
$$
D=\binom{r+1}{2}
$$
implies $\scs(\mathcal F_k)\ge D^kL$.
Therefore, since a valid relaxation sequence for $G_k$ must contain every element of $\cF_k$ as a subsequence, we have
\begin{equation}
\OPT_k \ge D^kL.
\label{eqn2}
\end{equation}

Combining the bounds (\ref{eqn1}) and (\ref{eqn2}) at depth $k$, we obtain that 
the gap ratio is greater than
$\frac12\left(\frac DA\right)^k$.

It remains to account for the construction size.
Let $n_k$ denote the number of vertices in $G_k$. We have $n_0 = L+1$, and $n_k=rn_{k-1}+\nu-2r$, for $k>0$.
In particular, denoting $C=L+\nu$, we have
$$n_k\le r^kC.$$

Define
$$
\gamma_r
=\log_r\!\left(\frac DA\right)
=\log_r\!\left(\frac{r(r+1)}{2(2r-1)}\right).
$$
Since $D/A>r/4$,
$$
\gamma_r>1-\log_r4,
\qquad\text{so}\qquad
\gamma_r\longrightarrow1.
$$
For fixed \(\varepsilon\in(0,1)\), choose a fixed \(r\) large enough that
$\gamma_r>1-\varepsilon$.

Next, choose a fixed constant $d$ satisfying
$$
d\gamma_r>(d+1)(1-\varepsilon),
$$
and set
$$
k=\left\lceil d\log_rC\right\rceil.
$$
The number of vertices in $G_k$ is $$
n_k\le rC^{d+1},
$$
where $r$ and $d$ depend only on $\varepsilon$, and the gap ratio exceeds
$$
\frac12 C^{d\gamma_r}>n_k^{1-\varepsilon}
$$
for sufficiently large inputs (instances with small $C$ are decided by brute force in constant time).

An $n_k^{1-\varepsilon}$-approximation of the shortest valid relaxation sequence would thus distinguish the NO-case upper bound $2A^{k}L$ and the YES-case lower bound of $D^kL$; a polynomial running time would therefore imply P=NP.

Again, we note that the hardness result holds even for very sparse graphs, since both the number of vertices ($n_k$) and the number of edges ($m_k)$ are of the form $r^k L(1+o(1))$.

\section{Conclusions}

We showed that the number of edge-relaxations in Yen's variant of the Bellman-Ford algorithm cannot be improved in a deterministic non-adaptive way by any constant factor. Our lower bound is shown for the complete directed graph as the input. Besides serving as a worst-case example, the complete graph is also natural as it captures \emph{full non-adaptivity}: If an algorithm is oblivious to both the weights and the \emph{structure} of the graph, then it must attempt to relax every possible edge, as if the input were a complete graph. 

In the randomized case, a gap between the $(\frac{1}{12} - o(1))n^3$ lower bound due to Eppstein and the $(\frac{1}{3} + o(1))n^3$ upper bound due to Bannister and Eppstein remains~\cite{Eppstein, Bannister}.
For incomplete \emph{dense} graphs, i.e., with the number of edges $m \geq n^{1+ {\varepsilon}}$, Eppstein shows that in the worst case, the $O(mn)$ bound cannot be improved by non-adaptive relaxation sequences (in this case, the algorithm knows the graph structure but not the weights). For sparse families, a gap between the $O(mn)$ upper bound and an $\Omega(mn/\log{n})$ lower bound remains~\cite{Eppstein}. 

On the complexity side we establish the hardness of verifying the validity of a relaxation sequence, and of finding the optimal relaxation sequence for a given graph, or approximating it to a better than linear factor, answering open questions from the literature. The approximation lower bound is essentially as strong as possible, since an $n$-approximation is easily attained by a round-robin relaxation sequence (after a simple preprocessing that deletes edges unreachable via simple paths from $s$). 

\paragraph{AI disclosure.} All results excluding \S\,\ref{sec4} and \S\,\ref{sec:amp} were obtained without AI use. The first version of the reduction of \S\,\ref{sec4} was found by interactively exploring the problem with ChatGPT 5.5. This was subsequently modified, extended to a hardness-of-approximation proof, and written up in the current form by the authors. The idea and first sketch of the gap amplification (\S\,\ref{sec:amp}) came from the authors, with subsequent AI use in working out the details and verifying the proofs. The writeup is due to the authors who take full responsibility for the content of the paper.

\newpage

\small
\bibliographystyle{alpha}
\bibliography{main}

\end{document}